**Comment on "Exchange bias-like phenomenon in SrRuO$_3$" by Pi et al (Appl. Phys. Lett. 88, 102502 (2006).**

Lior Klein

Deaprtment of Physics,  Bar-Ilan University, Ramat-Gan 52900, Israel

Pi et al report in their letter [1] that SrRuO$_3$ exhibits "exchange bias-like phenomenon". In our comment we would like to point out that the data presented in the letter is qualitatively different from what is known as exchange-bias behavior; furthermore, it is completely consistent with normal behavior of ferromagnetic materials.

Exchange bias [2] is a well known phenomenon encountered in systems containing interfaces between ferromagnetic and antiferromagnetic materials. The observation in ferromagnetic-antiferromagnetic systems exhibiting exchange-bias is a shift of the magnetization hysteresis loops along *the field axis* while no shift is observed along the magnetization axis. In a sense the system behaves as if in addition to the external field there is an internal field acting on the ferromagnetic material which arises from the interface between the ferromagnetic and antiferromagnetic material.

"The exchange bias-like" behavior reported by Pi et al is quite different. They report that after field cooling SrRuO$_3$ from above the Curie temperature there is a shift in the hysteresis loops; however, the shift is not along the field axis (as seen in exchange-bias systems) but along the magnetization axis. Moreover, there is no shift of the entire hysteresis loop (as seen in exchange-bias systems) but a downward displacement of the positive remanent magnetization combined with an upward displacement of the negative remanent magnetization. These are essential differences which indicate that the observed behavior has no relation to exchange-bias.

The observed shift is not surprising at all. Due to the irreversible nature of the magnetization process, *any* ferromagnetic material after being field cooled from above its Curie temperature, exhibits minor hysteresis loops whose center is shifted in the magnetization axis, when the field sweeping range in the hysteresis loop is small enough. The origin of the shift is quite trivial. A field cooled ferromagnet is mostly magnetized and if the sweeping field is not sufficiently large to completely reverse the magnetization, the center of the hysteresis loop will be shifted along the magnetization axis. Naturally, as the sweeping range increases, the shift is expected to decrease, as indeed is observed by Pi et al.

In passing, we want to point out that the authors attribute the alleged exchange-bias behavior to the fact that "SrRuO$_3$ is a famous spin-glass". While there were few reports with such claims, it is important to note that many reports (including my own) indicate that SrRuO$_3$ is an itinerant ferromagnet and not a spin-glass [3].